\documentclass{elsart}
\usepackage{epsfig,amssymb}
\journal{Physica A}

\begin{document}
\def\eeq{\end{equation}}
\def\beq{\begin{equation}}
\def\bea{\begin{eqnarray}}
\def\eea{\end{eqnarray}}
\begin{frontmatter}

\title{Sensitivity function and entropy increase rates for $z$-logistic 
map family at the edge of chaos}

\author{Ahmet Celikoglu and}
\author{Ugur Tirnakli\corauthref{cor}} 
\corauth[cor]{Corresponding author.}
\ead{ugur.tirnakli@ege.edu.tr}
\address{Department of Physics,
Faculty of Science, Ege University, 35100 Izmir, Turkey}

\begin{abstract}
It is well known that, for chaotic systems, the production of relevant 
entropy (Boltzmann-Gibbs) is always linear and the system has strong 
(exponential) sensitivity to initial conditions. In recent years, 
various numerical results indicate that basically the same type of 
behavior emerges at the edge of chaos if a specific generalization of 
the entropy and the exponential are used. In this work, we contribute 
to this scenario by numerically analysing some generalized nonextensive 
entropies and their related exponential definitions using $z$-logistic 
map family. We also corroborate our findings by testing them at accumulation 
points of different cycles. 
\end{abstract}

\begin{keyword}
Nonlinear dynamics \sep   non-extensivity

\PACS  05.45.-a  \sep 05.20.-y 

\end{keyword}
\end{frontmatter}

\date{\today}
\maketitle

\section{Introduction}
In nonlinear dynamics, it is well known that the chaotic systems 
has an exponential sensitivity to initial conditions, characterized 
by the sensitivity function (for one-dimensional case) 
\beq
\label{sens}
\xi (t) \equiv \lim_{\Delta x(0)\rightarrow 0} \frac{\Delta x(t)}{\Delta x(0)},
\eeq 
(where $\Delta x(t)$ is the distance, in phase space, between two copies at time $t$) 
to diverge as $\xi(t)= \exp(\lambda_1 t)$, where  $\lambda_1$ is the standard Lyapunov 
exponent \cite{hilborn}. 
On the other hand, for the marginal case, where  $\lambda_1=0$, the form 
of the sensitivity function could be a whole class of functions. One of the candidates 
is a power-law behavior, which can be characterized by an appropriate generalization 
of exponentials, namely $\xi(t)= \widetilde{\exp}(\lambda t)$, where $\lambda$ is the 
generalized Lyapunov exponent. Therefore, $\lambda > 0$ and $\lambda < 0$ cases 
correspond to weak sensitivity and weak insensitivity to initial conditions respectively. 
This, in fact, constitutes a unified framework since the generalized exponentials 
include the standard one as a special case for an appropriate choice of related 
parameter. 

The other concept that we focus in this work is the entropy production. 
For a chaotic system, the Kolmogorov-Sinai (KS) entropy $K_1$ is defined as the 
increase, per unit time, of the standard Boltzmann-Gibbs entropy and it is 
basically related to the standard Lyapunov exponents through the Pesin 
identity, which states that $K_1 = \lambda_1$ if $\lambda_1 > 0$ and $K_1=0$ 
otherwise. Here, it is worth mentioning that the KS entropy is basically 
defined in terms of a single trajectory in phase space, using a symbolic 
representation of the regions of a partitioned phase space (see \cite{hilborn}). 
However, it appears that, in most cases, this definition can be replaced by one 
based on an ensemble of initial conditions, which is the version we use herein. 
In the framework of this version, it has already been shown that the statistical 
definition of entropy production rate exhibits a close analogy to the 
production rate of thermal entropy and practically coincides with the KS entropy 
in chaotic systems \cite{latora}. In recent years, there have been efforts on 
extending this picture to dynamical systems at their marginal points (like chaos 
threshold) by using a generalized entropic form, which allows us to define a 
generalized KS entropy $K$. From this definition, one can also conjecture 
a Pesin-like identity as $K=\lambda$, which recovers both the chaotic and 
critical (i.e., chaos thresholds) cases since it includes standard Pesin 
identity as a special case \cite{TPZ}. 

As a whole, this unified framework has been numerically verified firstly for 
the logistic map \cite{TPZ,latora2} using the Tsallis entropy 
$S_q \equiv \left(1- \sum_{i=1}^W p_i^q\right)/\left(q-1\right)$ \cite{tsa88}, 
which grows linearly for a special value of entropic index $q$, which is 
$q_{sen}\simeq 0.24$; whereas the asymptotic power-law sensitivity to initial 
conditions is characterized by the generalized exponential 
$\widetilde{\exp}(x)=\exp_q(x)=\left[1+(1-q)x\right]^{1/(1-q)}$ with the 
same value of the entropic index.
After these works on the logistic map, numerical evidences supporting this 
framework came also from the studies of other low-dimensional dynamical 
systems, such as the $z$-logistic map family \cite{costa,ugur1}, the Henon 
map \cite{ugur2} and the asymmetric logistic map family \cite{ugur3}. 
Besides these numerical investigations, analytical treatment of the subject 
is also available recently in a series of paper by Baldovin and 
Robledo \cite{Robledo}.
Very recently, a similar analysis has been performed for the $z$-logistic 
map family, but this time, using ensemble-averaged initial conditions 
distributed uniformly over the entire available phase space \cite{ananos,ugur4}. 
The most important outcome of this analysis is another numerical verification 
of the coincidence of the entropic indices coming from the sensitivity function 
and entropy production rates (although with a different value $q_{sen}^{av}\simeq 0.36$), 
which consequently broadens the validity of the Pesin-like identity. 

Finally, here we should mention a recent work of Tonelli et al. \cite{tonelli}, 
where they demonstrate that the above-mentioned framework is even more general by 
making use of a two-parameter family of logarithms \cite{kani}

\beq
\widetilde{\log}(\xi) = \frac{\xi^{\alpha}-\xi^{-\beta}}{\alpha+\beta}
\eeq
where $\alpha$ ($\beta$) characterizes the large (small) argument asymptotic behavior. 
From this wide class, they analysed four interesting one-parameter cases, namely ; \\
\noindent (i)~the Tsallis logarithm \cite{tsa88} : $\alpha=1-q$ and $\beta=0$  \\
\noindent (ii)~the Abe logarithm \cite{abe} : $\alpha=1-q$ and $\beta=\alpha/(1+\alpha)$ \\
\noindent (iii)~the Kaniadakis logarithm \cite{kani2} : $\alpha=\beta=\kappa=1-q$ \\
\noindent (iv)~the $\gamma$ logarithm : $\alpha=1-q$ and $\beta=(1-q)/2$. \\
\noindent Their analysis consists of studying the sensitivity function and the 
entropy increase rates for the logistic map at the edge of chaos. Obviously, for the 
corresponding entropy in each case, one needs to use 

\beq
S(t)=\sum_{i=1}^W p_i(t) \widetilde{\log}\left(\frac{1}{p_i(t)}\right)
\eeq
from where the entropy production rates in time can be calculated. Their numerical 
results clearly verified that, for the logistic map, the relavant value of the 
entropic index is $q_{sen}^{av}\simeq 0.36$ not only for the Tsallis case but also 
for the others as well, and moreover that the Pesin-like identity is also present 
for all cases (with different numerical values for each case).

The aim of the present effort is to check the validity of the above-mentioned 
picture making use of the $z$-logistic map family. Moreover, the effect of different 
cycles on this validity is tested by analysing four distinct cycles. 

\section{The model and the procedure}
The model system that we use in our analysis is the $z$-logistic map family 
defined as 

\beq
x_{t+1} = 1 - a |x_t|^z \;\;\; 
\eeq
where $z>1; \,0<a \le 2;\, |x_t| \le 1; \, t=0,1,2,...$. It is easily seen that 
$z=2$ case corresponds to the standard logistic map.

Firstly we study the sensitivity to initial conditions at the edge of chaos using 
the sensitivity function given in Eq.(1). From its definition, for the calculation, we 
proceed with considering two initially very close points, which makes for example 
$\Delta x(0)=10^{-12}$ and then at each time step we numerically calculate the 
sensitivity function. To make an ensemble average, this procedure is repeated 
many times starting from different initial conditions all chosen randomly within 
the phase space and an average is taken over all values of $\widetilde{\log}(\xi)$. 
The special value of $\alpha$ (consequently, $q_{sen}^{av}$) is found as the value 
for which we obtain a linear time dependence of $\left<\widetilde{\log}(\xi)\right>(t)$. 
Finally, it is clear that the slope of this curve gives us the generalized Lyapunov 
exponent $\lambda$. 
In our analysis we concentrate on various values of $z$ for four different cycles 
at their accumulation points denoted by $a_c$. These critical values are given in Table 1.

 \begin{center}
\begin{table}
 \caption{The critical $a_c$ values of the $z$-logistic map family 
for all cycles used in this study.\medskip}
 \label{tab:Table 1}
 \begin{tabular}{|c|c|c|c|c|}
  \hline
  $z$  &    \multicolumn{4}{c|}{\em $a_c$}  \\
  \cline{2-5}
    &   cycle 2  & cycle 3 & cycle 4 &  cycle 5 \\
  \hline
  1.75  & 1.35506...  & 1.74730...  & 1.92764... & 1.60749... \\
  2     & 1.40115...  & 1.77981...  & 1.94217... & 1.63101... \\
  2.5   & 1.47054...  & 1.82886...  & 1.96144... & 1.66954... \\
  3     & 1.52187...  & 1.86299...  & 1.97302... & 1.69944... \\
  4     & 1.59490...  & 1.90597...  & 1.98524... & 1.74282... \\
 \hline 
\end{tabular}
\end{table}
\end{center}

Secondly we study the entropy production rates of the entropies related to four  
selected logarithms, namely we use Eq.(3).
The procedure that we implement for the entropy production is the same as the one 
introduced firstly in \cite{latora} for chaotic systems and then used in other 
works related to this context. It consists of dividing the phase space in $W$ 
equal intervals and putting randomly $N$ initial points in one of them. Then, 
one should trace the spread of initial points within the phase space and 
calculate the entropy $S(t)$, from where we can obtain the entropy production 
per unit time as 

\beq
K = \lim_{t\rightarrow\infty} \lim_{W\rightarrow\infty}
\lim_{N\rightarrow\infty} \frac{S(t)}{t} .
\eeq
We then repeat this many times starting from randomly chosen different intervals 
of the phase space and average the entropy $S(t)$ over these experiments. 
Here, the special value of $\alpha$ (consequently, $q_{sen}^{av}$) is obviously 
found as the value for which we obtain a linear time dependence of $\left<S\right>(t)$. 
In addition to this, the value of $K$ can be obtained from the slope of this 
linear curve.

From the previous results \cite{ananos,ugur4}, it is known for the Tsallis case that, 
for the $z$-logistic map family, the proper $\alpha$ values (namely, $q_{sen}^{av}$) 
coming from the sensitivity function and entropy production coincide each other and 
moreover, for all $z$ values, for the proper $\alpha$ value the Pesin-like identity 
is preserved as $K=\lambda$. Then, very recently, for the standard logistic map 
(namely, $z=2$ case), Tonelli et al. \cite{tonelli} showed that the picture is 
wider that this by considering the other three cases mentioned above. Now, our aim is 
to check the validity of this framework for the $z$-logistic map and for various 
cycles.  

\section{Numerical results and conclusions}
In our simulations, for the sensitivity function analysis, for each cycle and $z$ 
value we consider two very close points so that $\Delta x(0)=10^{-12}$ and then 
numerically calculate the sensitivity function $\xi$ from its definition. 
In order to take the logarithm of $\xi$, we use four different cases mentioned 
previously. We repeat this for $4\times 10^7$ uniformly distributed initial 
conditions to make an ensemble average of $\left<\widetilde{\log}(\xi)\right>(t)$. 
Its time evolution is given for a representative value of $z$ for cycles 3, 4 and 
5 in Fig.1(a),(c) and (e) respectively. For each case (namely, Tsallis, Abe, 
Kaniadakis and $\gamma$ cases), the special value of $\alpha$ is 
determined by testing various $\alpha$ values until we obtain a linear time 
dependence. From the slope of each curve, the corresponding generalized Lyapunov 
exponents $\lambda$ can be calculated. All of the obtained results for 
$\alpha$ and $\lambda$ are given in Table 2. For a given $z$ value of each 
cycle, it is seen that the special value of $\alpha$ is the same (within error 
bars) for all four generalized cases. However, generically the generalized 
Lyapunov exponents are different in each case.

Then, we study the entropy increase rates. We divide the phase space into $W$ 
equal cells and put $N$ initial conditions into a randomly chosen cell. 
We then let the dynamics evolve in time and calculate the generalized 
entropies. In order to make an ensemble average, we repeat this procedure 
many times starting from randomly chosen different cells ($W/2$ cells), 
which reduces also the fluctuations that appear at the edge of chaos.
In all our simulations we use $W=10^5$ and $N=10 W$. For each entropic 
form, we use the special value of $\alpha$ calculated from the sensitivity 
function and we obtain the linear coefficient $K$ from the slope of the 
linear entropy increase rates as it is seen in Fig.1(b),(d) and (f).  
As it is clear from Table 2 that, for all $z$ values of each cycle and for 
all entropic forms, $K$ values coincide with corresponding $\lambda$ values 
(within numerical errors).

\begin{figure*}[htb]
 \begin{minipage}[htb]{0.5\textwidth}
 \begin{center}
  \epsfig{file=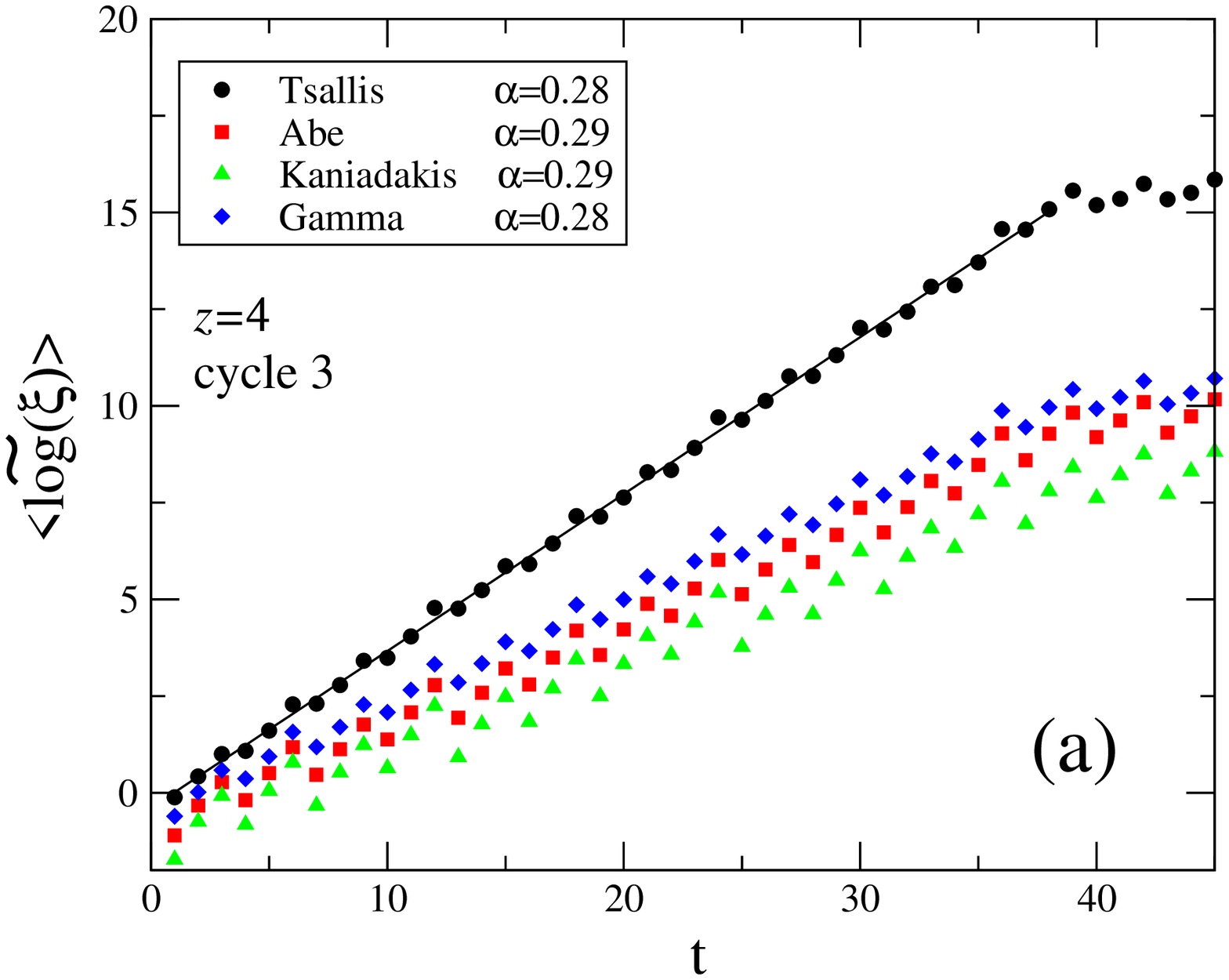,width=\textwidth,clip=}
 \end{center}
 \end{minipage}
 \begin{minipage}[htb]{0.5\textwidth}
 \begin{center}
  \epsfig{file=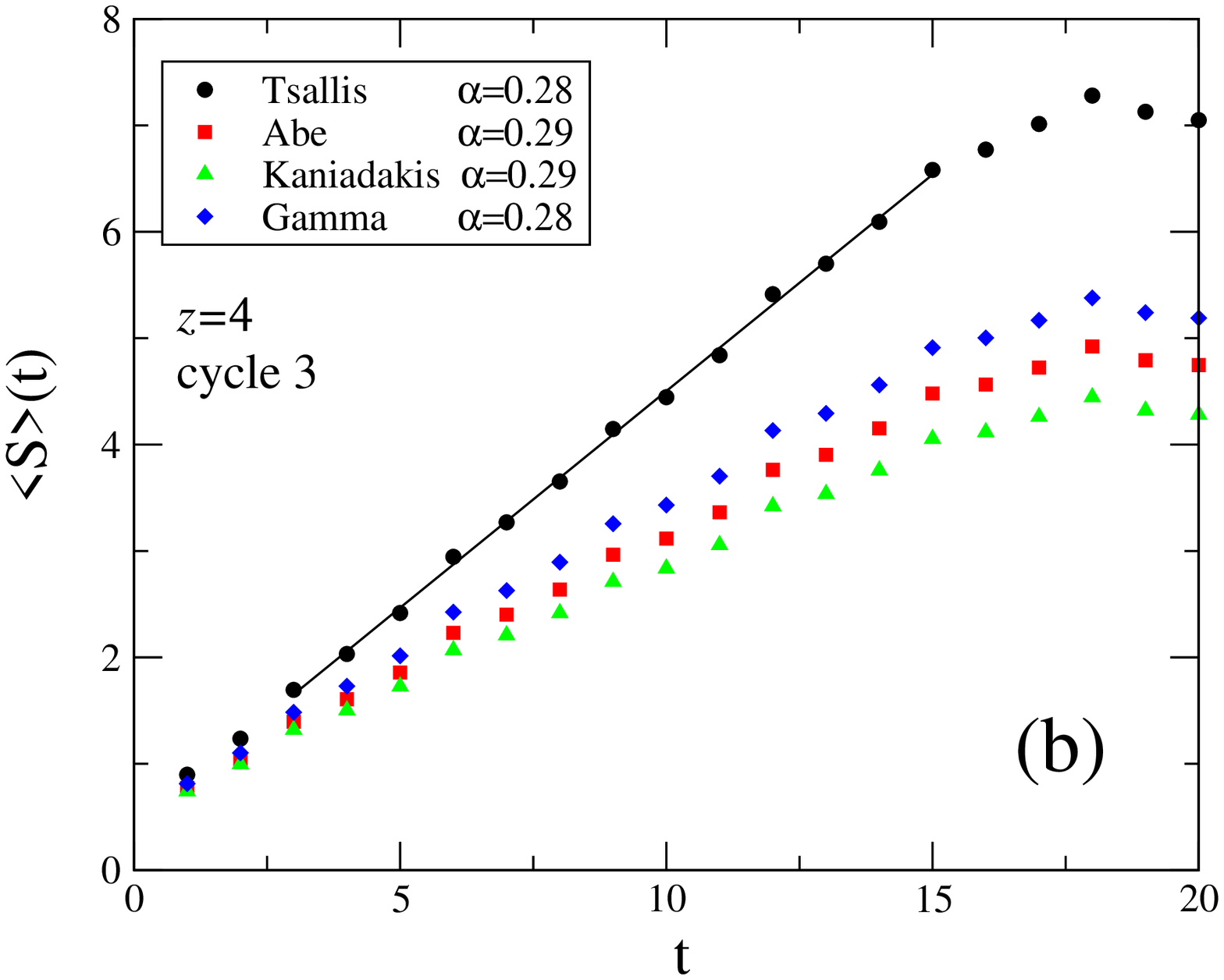,width=\textwidth,clip=}
 \end{center}
 \end{minipage}
\end{figure*}

\begin{figure*}[htb]
 \begin{minipage}[htb]{0.5\textwidth}
 \begin{center}
  \epsfig{file=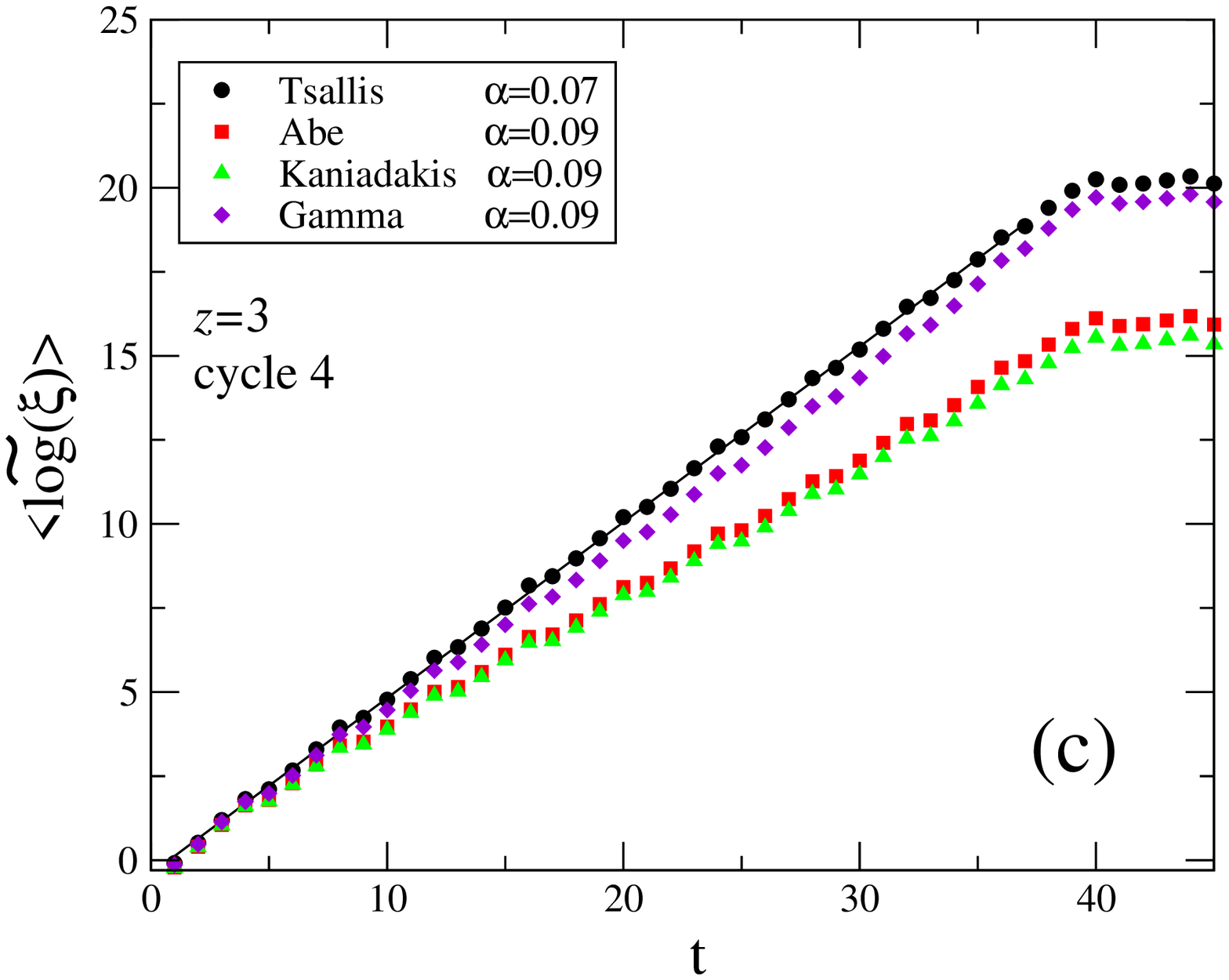,width=\textwidth,clip=}
 \end{center}
 \end{minipage}
 \begin{minipage}[htb]{0.5\textwidth}
 \begin{center}
  \epsfig{file=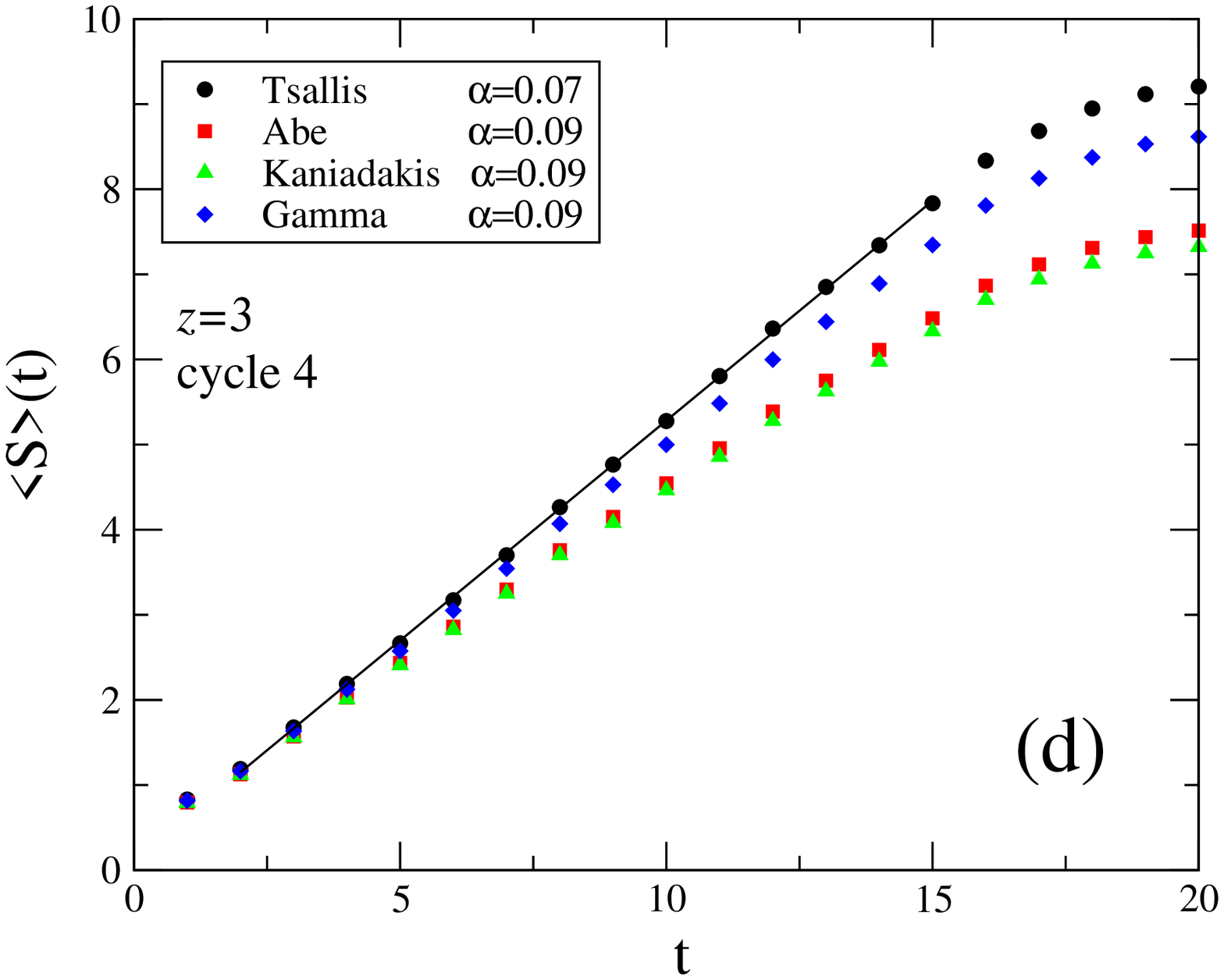,width=\textwidth,clip=}
 \end{center}
 \end{minipage}
\end{figure*}

\begin{figure}[htb]
 \begin{minipage}[htb]{0.5\textwidth}
 \begin{center}
  \epsfig{file=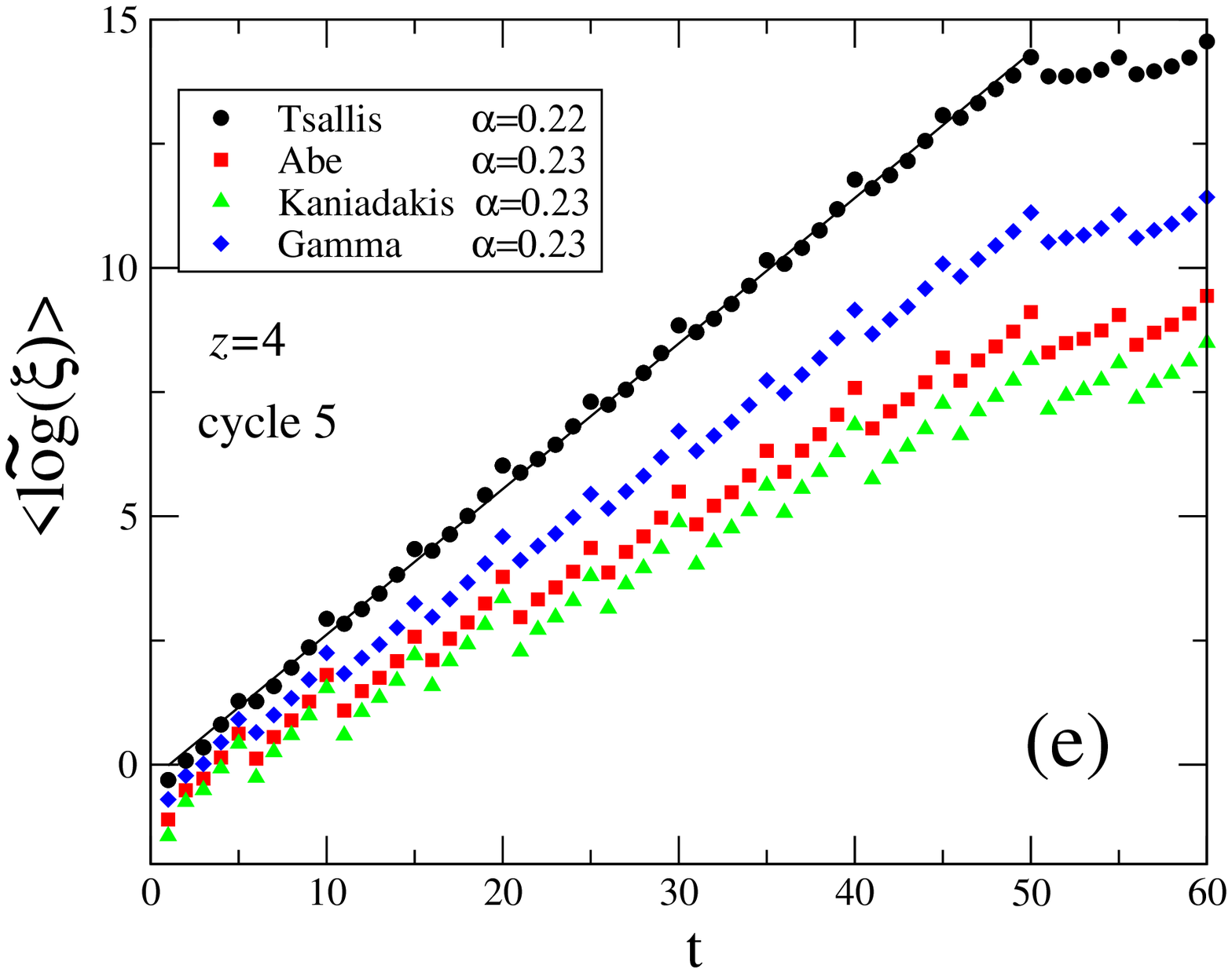,width=\textwidth,clip=}
 \end{center}
 \end{minipage}
 \begin{minipage}[htb]{0.5\textwidth}
 \begin{center}
  \epsfig{file=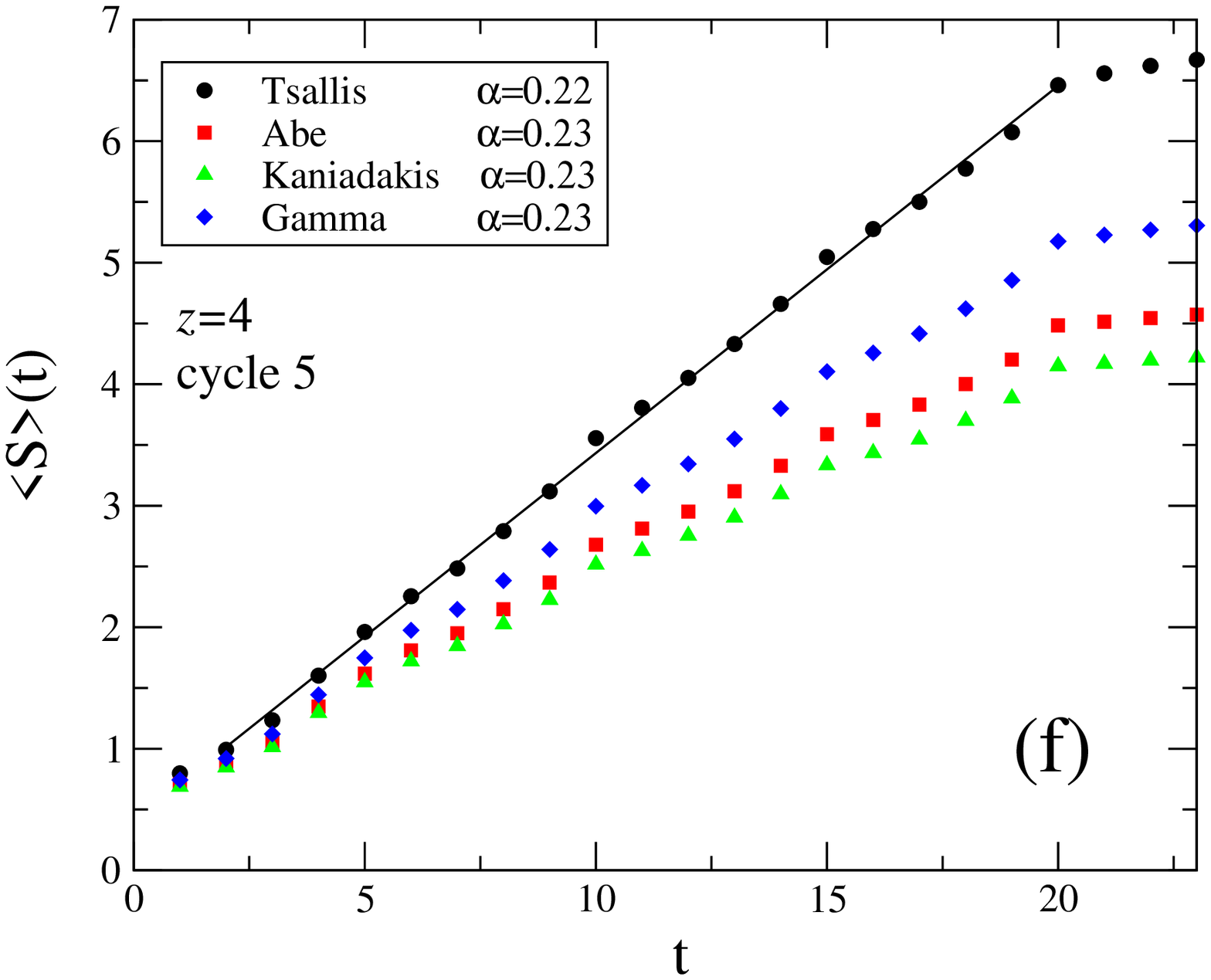,width=\textwidth,clip=}
 \end{center}
 \end{minipage}
 \caption{Time evolution of the sensitivity function averaged over 
$4\times 10^7$ uniformly distributed initial conditions for (a)~cycle 3 and $z=4$, 
(c)~cycle 4 and $z=4$, (e)~cycle 5 and $z=4$. Entropy increase rates with 
$W=10^5$ and $N=10 W$ are given in (b),(d) and (f) for the same cycle and $z$ values 
chosen for the sensitivity function. In each case an average over $W/2$ boxes are performed.}
 \label{fig:D-vs-time}
\end{figure}

\begin{figure}[htb]
 \begin{center}
  \epsfig{file=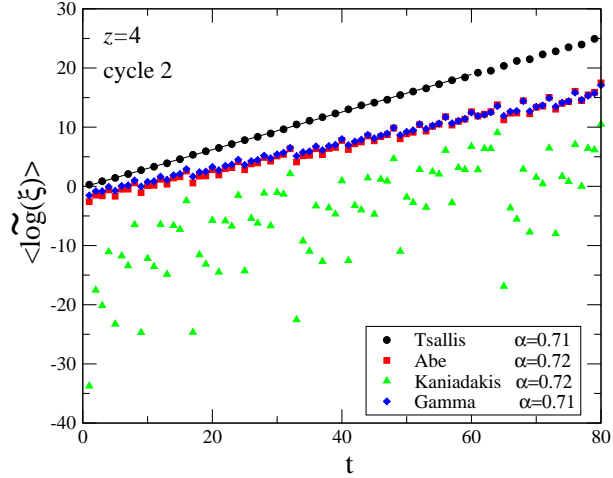,width=8cm}
 \end{center}
%
 \caption{(a) Time evolution of the sensitivity function averaged over 
$4\times 10^7$ uniformly distributed initial conditions for cycle 2 and $z=4$.}
\end{figure}

Finally, it is worth mentioning that generically for the increasing values of $z$ 
(depending also on the cycle), the observed fluctuations in the time evolution of 
the sensitivity function become to be more stronger. This is even very much stronger 
in Kaniadakis case than other three cases. Due to these fluctuations, we did not 
give any value for $z=3,4$ cases of cycle 2 in Table 2. A visualization of these  
fluctuations is given in Fig.~2. The reason for this is the following : If a given 
initial condition has weak insensitivity then $\xi$ value is smaller than 1, which 
makes the value of each generalized logarithm negative. But, whenever the $\xi$ 
value is closer to 0, then logarithms start to go very large negative values and the 
most rapidly diverging case is the Kaniadakis one. As a result, these very large negative 
values dominate the sum for the average which yield large fluctuations.

Summing up, in this work, we have numerically analysed the sensitivity to initial 
conditions and entropy increase rates of $z$-logistic map family at the edge of 
chaos for various cycles. We have discussed the same generalized logarithms and 
their related entropies studied in \cite{tonelli} for the standard logistic map and 
shown that the framework is similar to that of the standard logistic map; namely, 
(i)~the proper $\alpha$ value, which gives a linear time dependence of the 
sensitivity function and a linear entropy production, is the same for all four 
entropic forms for a given $z$ value of a given cycle, 
(ii)~the results corroborate a generalized Pesin-like identity $K=\lambda$ for each 
entropic form with a different numerical value.

\begin{center}
\begin{table}
\caption{For the generalized logarithms given in the text, three important quantities 
are listed for cycles 2, 3, 4 and 5 : (i)~the appropriate $\alpha$ value which gives a 
linear time dependence of $\left<\widetilde{\log}(\xi)\right>(t)$, 
(ii)~the value of the generalized Lyapunov exponent $\lambda$ which is the slope of 
the above-mentioned linear time dependence and 
(iii)~the value of entropy increase rates $K$ which is calculated from the slope of the 
linearly increased entropies coming from each of the generalized logarithms.
For many cases, the error bars are $\pm 0.01$ and in a few cases, they are found as 
$\pm 0.02$.}
\label{tab:Table2}
\begin{tabular}{||c|c|c|c|c|c|c|c|c|c|c|c|c|c||} \hline\hline
cycle & $z$  & \multicolumn{3}{c|} {Tsallis} & \multicolumn{3}{c|} {Abe} & \multicolumn{3}{c|} {Kaniadakis} & \multicolumn{3}{c||} {$\gamma$} \\ \cline{3-14}
      &      & $\alpha$ & $K$  & $\lambda$ & $\alpha$ &  $K$ & $\lambda$ & $\alpha$ &  $K$ & $\lambda$ & $\alpha$ &  $K$ & $\lambda$ \\ \hline\hline
      &  2   & 0.64 & 0.26  & 0.27 & 0.65 & 0.17  & 0.18 & 0.65 & 0.14 & 0.15 & 0.65 & 0.19  & 0.19 \\ \cline{2-14}
  2   &  2.5 & 0.66 & 0.28  & 0.28 & 0.67 & 0.19  & 0.19 & 0.69 & 0.15 & 0.18 & 0.67 & 0.20  & 0.20 \\ \cline{2-14}
      &  3   & 0.67 & 0.28  & 0.28 & 0.67 & 0.18  & 0.18 & $--$ & $--$  & $--$ & 0.67 &  0.19 & 0.19 \\ \cline{2-14}
      &  4   & 0.71 & 0.31  & 0.31 & 0.72 & 0.21  & 0.21 & $--$ & $--$  & $--$ & 0.71 & 0.21  & 0.21 \\ \cline{1-14}\hline\hline
      & 1.75 & 0.08 & 0.47  & 0.48 & 0.10 & 0.37  & 0.37 & 0.10 & 0.36 & 0.35 & 0.09 & 0.42 & 0.41 \\ \cline{2-14}
      &  2   & 0.12 & 0.48  & 0.49 & 0.14 & 0.35  & 0.36 & 0.14 & 0.33 & 0.34 & 0.13 & 0.40 & 0.40 \\ \cline{2-14}
  3   &  2.5 & 0.18 & 0.47  & 0.48 & 0.19 & 0.30  & 0.31 & 0.19 & 0.28 & 0.29 & 0.19 & 0.37 & 0.38 \\ \cline{2-14}
      &  3   & 0.22 & 0.44  & 0.45 & 0.23 & 0.29  & 0.29 & 0.23 & 0.26 & 0.26 & 0.23 & 0.35 & 0.35 \\ \cline{2-14}
      &  4   & 0.28 & 0.41  & 0.41 & 0.29 & 0.26  & 0.26 & 0.29 & 0.23 & 0.23 & 0.28 & 0.28 & 0.28 \\ \cline{1-14}\hline\hline
      & 1.75 & 0.01 & 0.58  & 0.59 & 0.03 & 0.56  & 0.57 & 0.03 & 0.56 & 0.56 & 0.02 & 0.58 & 0.60 \\ \cline{2-14}
      &  2   & 0.02 & 0.57  & 0.57 & 0.05 & 0.53  & 0.55 & 0.05 & 0.53 & 0.54 & 0.03 & 0.55 & 0.56 \\ \cline{2-14}
  4   &  2.5 & 0.05 & 0.56  & 0.58 & 0.07 & 0.47  & 0.47 & 0.07 & 0.46 & 0.45 & 0.06 & 0.51 & 0.52 \\ \cline{2-14}
      &  3   & 0.07 & 0.52  & 0.53 & 0.09 & 0.41  & 0.41 & 0.09 & 0.40 & 0.39 & 0.09 & 0.48 & 0.50 \\ \cline{2-14}
      &  4   & 0.11 & 0.46  & 0.46 & 0.13 & 0.34  & 0.34 & 0.13 & 0.32 & 0.32 & 0.13 & 0.40 & 0.42 \\ \cline{1-14}\hline\hline
      & 1.75 & 0.04 & 0.40  & 0.41 & 0.06 & 0.35  & 0.35 & 0.06 & 0.34 & 0.34 & 0.05 & 0.37 & 0.38 \\ \cline{2-14}
      &  2   & 0.07 & 0.39  & 0.42 & 0.09 & 0.32  & 0.32 & 0.09 & 0.31 & 0.31 & 0.08 & 0.35 & 0.36 \\ \cline{2-14}
  5   &  2.5 & 0.12 & 0.37  & 0.38 & 0.14 & 0.28  & 0.28 & 0.14 & 0.26 & 0.26 & 0.13 & 0.31 & 0.31 \\ \cline{2-14}
      &  3   & 0.16 & 0.35  & 0.35 & 0.18 & 0.25  & 0.25 & 0.18 & 0.23 & 0.24 & 0.17 & 0.28 & 0.28 \\ \cline{2-14}
      &  4   & 0.22 & 0.30  & 0.30 & 0.23 & 0.19  & 0.19 & 0.23 & 0.18 & 0.18 & 0.23 & 0.23 & 0.23 \\ \cline{1-14}\hline\hline
\end{tabular}
\end{table}
\end{center}

\section{Acknowledgements}
One of us (UT) would really like to express his acknowledgements to 
A. Robledo for his continual constructive comments and encouragement on the subject 
throughout many years. 
This work is supported by TUBITAK (Turkish agency) under the research project 104T148.


\end{document}